\documentclass[twocolumn, twocolappendix]{aastex631}
\usepackage{natbib}
\usepackage{latexsym}
\usepackage{graphicx}
\usepackage{epsfig}
\usepackage{amssymb}
\usepackage{amsmath}
\usepackage{epstopdf}
\usepackage{hyperref}
\usepackage{xcolor}

\newcommand{\yL}{\ensuremath{\mathcal{Y}_{\rm{L}}}}
\newcommand{\yS}{\ensuremath{\mathcal{Y}_{\rm{S}}}}
\newcommand{\justgrad}{\ensuremath{\nabla}}
\newcommand{\gradrad}{\ensuremath{\nabla_{\rm{rad}}}}
\newcommand{\gradad}{\ensuremath{\nabla_{\rm{ad}}}}

\newcommand{\gradmu}{\ensuremath{\nabla_{\mu}}}

\newcommand{\Ro}{\ensuremath{\mathrm{R}_{\rho}}}

\newcommand{\mS}{\ensuremath{\mathcal{S}}}
\newcommand\Pran{\ensuremath{\mathrm{Pr}}}
\newcommand{\brunt}{{Brunt--V\"{a}is\"{a}l\"{a}}}

\renewcommand{\vec}[1]{\mathbf{#1}}
\newcommand{\M}[1]{\mathbf{#1}}
\renewcommand{\dot}{\vec{\cdot}}
\renewcommand{\bar}[1]{\overline{#1}}
\newcommand{\grad}{\vec{\nabla}}

\newcommand{\editone}[1]{\textcolor{black}{#1}}

\received{}
\revised{}
\accepted{}
\published{}
\submitjournal{ApJ}

\shorttitle{Schwarzschild or Ledoux}
\shortauthors{Anders et al}

\begin{document}

\title{Schwarzschild and Ledoux are equivalent on evolutionary timescales}
\author[0000-0002-3433-4733]{Evan H. Anders}
\affiliation{CIERA, Northwestern University, Evanston IL 60201, USA}
\affiliation{Kavli Institute for Theoretical Physics, University of California, Santa Barbara, CA 93106, USA}
\author[0000-0001-5048-9973]{Adam S. Jermyn}
\affiliation{Center for Computational Astrophysics, Flatiron Institute, New York, NY 10010, USA}
\affiliation{Kavli Institute for Theoretical Physics, University of California, Santa Barbara, CA 93106, USA}
\author[0000-0002-7635-9728]{Daniel Lecoanet}
\affiliation{CIERA, Northwestern University, Evanston IL 60201, USA}
\affiliation{Department of Engineering Sciences and Applied Mathematics, Northwestern University, Evanston IL 60208, USA}
\affiliation{Kavli Institute for Theoretical Physics, University of California, Santa Barbara, CA 93106, USA}
\author[0000-0003-4323-2082]{Adrian E. Fraser}
\affiliation{Department of Applied Mathematics, Baskin School of Engineering, University of California, Santa Cruz, CA 95064, USA}
\affiliation{Kavli Institute for Theoretical Physics, University of California, Santa Barbara, CA 93106, USA}
\author[0000-0002-4538-7320]{Imogen G. Cresswell}
\affiliation{Department Astrophysical and Planetary Sciences \& LASP, University of Colorado, Boulder, CO 80309, USA}
\affiliation{Kavli Institute for Theoretical Physics, University of California, Santa Barbara, CA 93106, USA}
\author[0000-0002-8717-127X]{Meridith Joyce}
\affiliation{Space Telescope Science Institute, 3700 San Martin Drive, Baltimore, MD 21218, USA}
\affiliation{Kavli Institute for Theoretical Physics, University of California, Santa Barbara, CA 93106, USA}
\author[0000-0003-2124-9764]{J. R. Fuentes}
\affiliation{Department of Physics and McGill Space Institute, McGill University, 3600 rue University, Montreal, QC H3A 2T8, Canada}

\correspondingauthor{Evan H. Anders}
\email{evan.anders@northwestern.edu}

\begin{abstract}
    Stellar evolution models calculate convective boundaries using either the Schwarzschild or Ledoux criterion, but confusion remains regarding which criterion to use.
    Here we present a 3D hydrodynamical simulation of a convection zone and adjacent radiative zone, including both thermal and compositional buoyancy forces.
    As expected, regions which are unstable according to the Ledoux criterion are convective.
    Initially, the radiative zone adjacent to the convection zone is Schwarzschild-unstable but Ledoux-stable due to a composition gradient.
    Over many convective overturn timescales the convection zone grows via entrainment.
    The convection zone saturates at the size originally predicted by the Schwarzschild criterion, although in this final state the Schwarzschild and Ledoux criteria agree.
    Therefore, the Schwarzschild criterion should be used to determine the size of stellar convection zones, except possibly during short-lived evolutionary stages in which entrainment persists.
\end{abstract}


\section{Introduction}
\label{sec:introduction}
The treatment of convective boundaries in stars is a long-standing problem in modern astrophysics.
Models and observations disagree about the sizes of convective cores \citep{claret_torres_2018, joyce_chaboyer_2018b, viani_basu_2020, pedersen_etal_2021, johnston_2021}, the depths of convective envelopes \citep[inferred from lithium abundances;][]{pinsonneault_1997, sestito_randich_2005, carlos_etal_2019, dumont_etal_2021}, and the sound speed at the base of the Sun's convection zone \citep[see][Sec.~7.2.1]{basu_2016}.
Inaccurate convective boundary specification can have astrophysical impacts by e.g., affecting mass predictions of stellar remnants \citep{farmer_etal_2019, mehta_etal_2022} and the inferred radii of exoplanets \citep{basu_etal_2012, morrell_2020}.

In order to resolve the many uncertainties involved in treating convective boundaries, it is first crucial to determine the boundary location.
Some stellar evolution models determine the location of the convection zone boundary using the \emph{Schwarzschild criterion}, by comparing the radiative and adiabatic temperature gradients.
In other models, the convection zone boundary is determined by using the \emph{Ledoux criterion}, which also accounts for compositional stratification \citep[][chapter 3, reviews these criteria]{salaris_cassisi_2017}.
Recent work states that these criteria \emph{should} agree on the location of the convective boundary \citep{gabriel_etal_2014, mesa4, mesa5}, but in practice they can disagree \citep[see][chapter 2]{kaiser_etal_2020} \editone{which can affect asteroseismic observations \citep{silva_aguirre_etal_2011}.}
\editone{Efforts to properly choose convective boundaries locations have produced} a variety of algorithms in stellar evolution software instruments \citep{mesa4,mesa5}.

Multi-dimensional simulations can provide insight into the treatment of convective boundaries.
Such simulations show that a convection zone adjacent to a Ledoux-stable region can expand by entraining material from the stable region \citep{meakin_arnett_2007, woodward_etal_2015, jones_etal_2017, cristini_etal_2019, fuentes_cumming_2020, andrassy_etal_2020, andrassy_etal_2021}.
However, past simulations have not achieved a statistically-stationary state, leading to uncertainty in how to include entrainment in 1D models \citep{staritsin_2013, scott_etal_2021}.

In this letter, we present a 3D hydrodynamical simulation with a convection zone that is adjacent to a Ledoux-stable but Schwarzschild-unstable region.
Convection entrains material until the adjacent region is stable by both criteria.
Our simulation demonstrates that the Ledoux criterion \emph{instantaneously} describes the size of a convection zone.
However, when the Ledoux and Schwarzschild criteria disagree, the Schwarzschild criterion correctly predicts the size at which a convection zone saturates.
Therefore, when evolutionary timescales are much larger than the convective overturn timescale \citep[e.g., on the main sequence;][]{georgy_etal_2021}, the Schwarzschild criterion properly predicts convective boundary locations.
When correctly implemented, the Ledoux criterion should return the same result \citep{gabriel_etal_2014}.
We discuss these criteria in Sec.~\ref{sec:theory}, describe our simulation in Sec.~\ref{sec:results}, and briefly discuss the implications of our results for 1D stellar evolution models in Sec.~\ref{sec:conclusions}.

\begin{figure*}[t!]
\centering
\includegraphics[width=\textwidth]{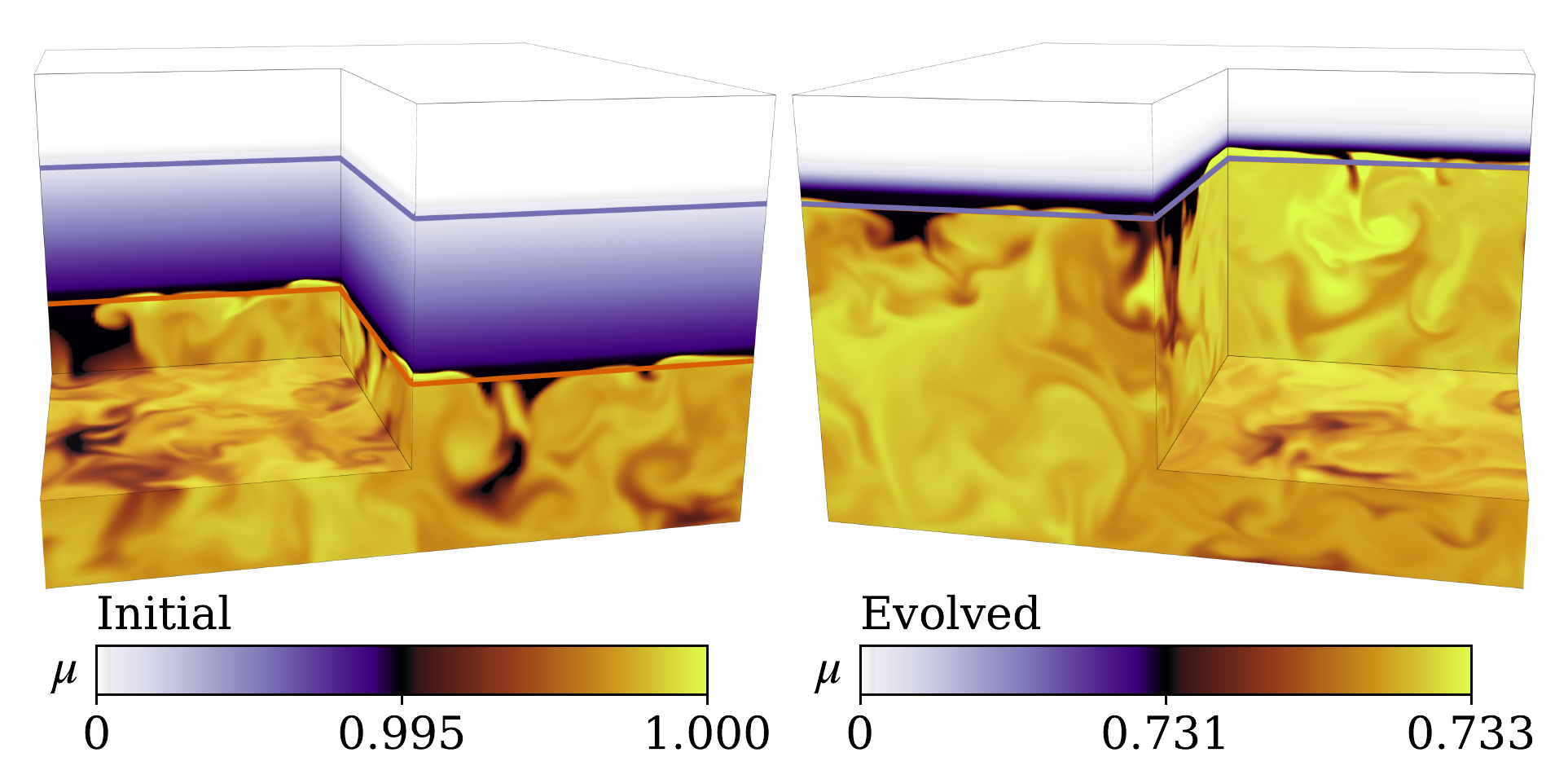}
\caption{
    Volume renderings of the composition $\mu$ at early (left) and late (right) times.
    A stable composition gradient is denoted by the changing color from top of the box (white) to the top of the convection zone (dark purple).
    The convection zone is well-mixed, so we expand the colorbar scaling there; black low-$\mu$ fluid is mixed into the yellow high-$\mu$ convection zone.
    Orange and purple horizontal lines respectively denote the heights at which $\yL = 0$ and $\yS = 0$.
    The Schwarzschild and Ledoux criteria are equivalent in the right panel, so the orange line is not visible.
    The simulation domain spans $z \in [0, 3]$, but we only plot $z \in [0, 2.5]$ here.
    A movie version of this figure is available \editone{online in the HTML version of the paper and} in the supplementary materials \citep{supp}\editone{; in the movie version, the initial Ledoux boundary height is denoted as a dotted orange line}.
\label{fig:dynamics}
}
\end{figure*}

\section{Theory \& Experiment}
\label{sec:theory}
The Schwarzschild criterion for convective stability is
\begin{equation}
    \yS \equiv \gradrad - \gradad < 0,
    \label{eqn:yS}
\end{equation}
whereas the Ledoux criterion for convective stability is
\begin{equation}
    \yL \equiv \yS +  \frac{\chi_\mu}{\chi_T}\gradmu < 0.
    \label{eqn:yL}
\end{equation}
The temperature gradient $\justgrad \equiv d \ln P / d \ln T$ (pressure $P$ and temperature $T$) is $\gradad$ for an adiabatic stratification and $\gradrad$ if all the flux is carried radiatively.
The Ledoux criterion includes the effects of the composition gradient $\gradmu = d\ln\mu/d\ln P$ (mean molecular weight $\mu$), where $\chi_T = (d\ln P / d\ln T)_{\rho,\mu}$ and $\chi_\mu = (d\ln P / d\ln\mu)_{\rho,T}$ (density $\rho$).

Stellar structure software instruments assume that convective boundaries coincide with sign changes of $\yL$ or $\yS$ \citep[][sec.~2]{mesa4}.
The various stability regimes that can occur in stars are described in section 3 and figure 3 of \citet{salaris_cassisi_2017}, but we note four important regimes here:
\begin{enumerate}
    \item Convection Zones (CZs): Regions with ${\yL > 0}$ are convectively unstable.
    \item Radiative Zones (RZs): Regions with $\yL \leq \yS < 0$ are always stable to convection.
        Other combinations of $\yL$ and $\yS$ \emph{may} also be stable RZs, as detailed below in \#3 and \#4.
    \item ``Semiconvection'' Zones (SZs): Regions with ${\yS > 0}$ but $\yL < 0$ are stablized by a composition gradient despite an unstable thermal stratification.
        These regions can be stable RZs or linearly unstable to oscillatory double-diffusive convection \citep[ODDC, see][chapters 2 and 4]{garaud_2018}.
    \item ``Thermohaline'' Zones: A stable thermal stratification can overcome an unstable composition gradient in regions with $\yS < \yL < 0$.
        These regions can be stable RZs or linearly unstable to thermohaline mixing \citep[see][chapters 2 and 3]{garaud_2018}.
\end{enumerate}
In this letter, we study a three-layer 3D simulation of convection.
The initial structure of the simulation is an unstable CZ (bottom, \#1), a compositionally-stabilized SZ (middle, \#3), and a thermally stable RZ (top, \#2).
We examine how the boundary of the CZ evolves through entrainment.
In particular, we are interested in determining whether the heights at which $\yS = 0$ and $\yL = 0$ coincide on timescales that are long compared to the convective overturn timescale.

Our simulation uses the Boussinesq approximation, which is formally valid when motions occur on length scales much smaller than the pressure scale height.
This approximation fully captures nonlinear advective mixing near the CZ-SZ boundary, which is our primary focus.
Our simulations use a height-dependent $\gradrad$ and buoyancy is determined by a combination of the composition and the temperature stratification, so $\yS$ and $\yL$ are determined independently and self-consistently.
Our simulation length scales are formally much smaller than a scale height, but a useful heuristic is to think of our 3D convection zone depth (initially 1/3 of the simulation domain) as being analogous to the mixing length in a 1D stellar evolution model.
For details on our model setup and Dedalus \citep{burns_etal_2020} simulations, we refer the reader to appendices \ref{app:model} and \ref{app:simulation_details}.

\editone{
    While $\mu$ represents the mean molecular weight in stellar modeling (e.g., Eqn.~\ref{eqn:yL}), throughout the rest of this manuscript we will use $\mu$ to denote the composition field in our dynamical model. 
    In stellar modeling, the quantity that determines convective stability \citep[the $B$ term in e.g.,][]{unno_etal_1989, mesa2} is obtained by accounting for the variation of pressure with composition in the full equation of state. 
    In our simulation, we employ an ideal equation of state in which compositional stability is determined by the gradient of $\mu$.
}

\begin{figure*}[t]
\centering
\includegraphics[width=\textwidth]{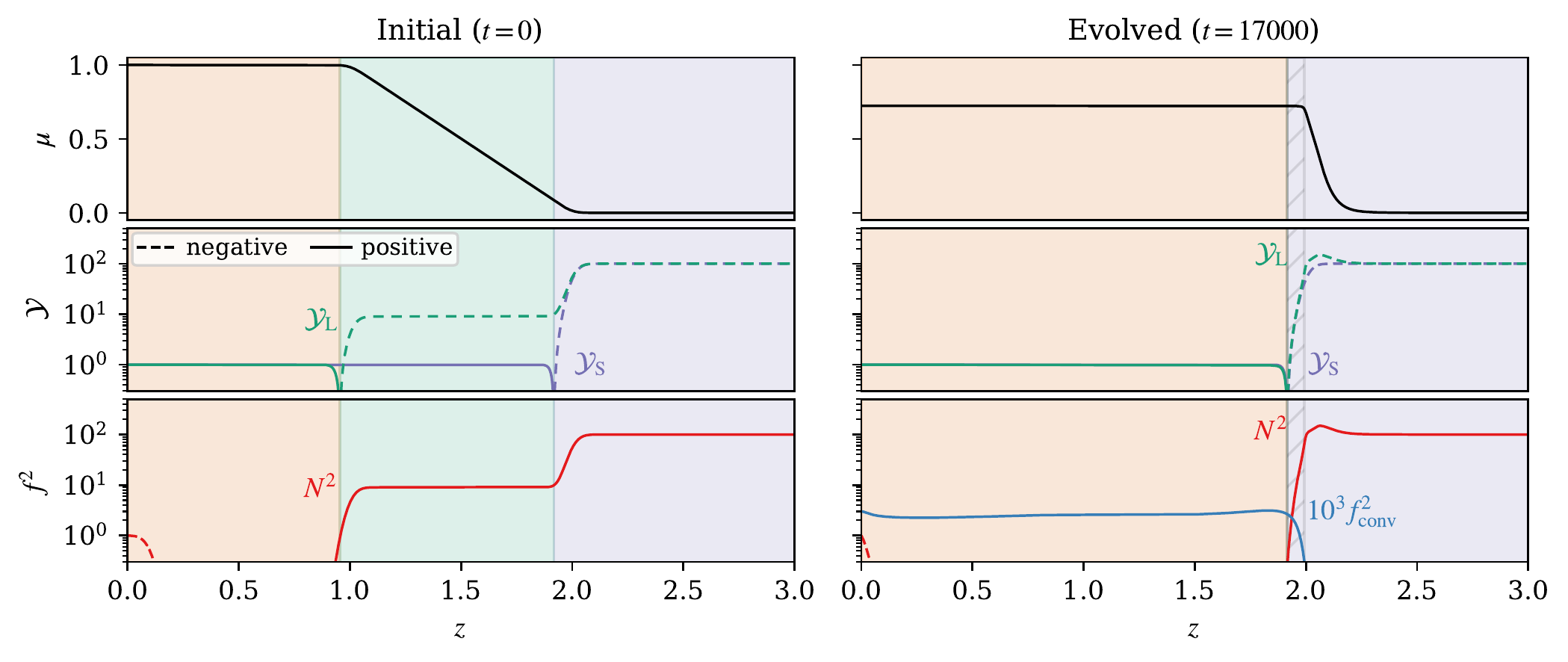}
\caption{
    Horizontally-averaged profiles of the composition (top), the discriminants $\yS$ and $\yL$ (middle, Eqns.~\ref{eqn:yS} \& \ref{eqn:yL}), and important frequencies (bottom, the \brunt$\,$ frequency $N^2 = -\yL$ and the square convective frequency $f_{\rm{conv}}^2$, see Eqn.~\ref{eqn:fconv2}).
    Positive and negative values are respectively solid and dashed lines.
    We show the initial (left) and evolved (right, time-averaged over 100 convective overturn times) states.
    There are no motions in the initial state, so $f_{\rm{conv}}^2 = 0$ and does not appear.
    The background color is orange in the CZ, green in the SZ, and purple in the RZ.
    The lightly hashed background region in the evolved RZ is the mechanical overshoot zone.
\label{fig:profiles}
}
\end{figure*}

\section{Results}
\label{sec:results}

In Fig.~\ref{fig:dynamics}, we visualize the composition field in our simulation near the initial state (left) and evolved state (right).
Thick horizontal lines denote the convective boundaries per the Ledoux (orange, $\yL = 0$) and Schwarzschild (purple, $\yS=0$) criteria.
Initially, the bottom third of the domain is a CZ, the middle third is an SZ, and the top third is an RZ.
Convection motions extend beyond $\yL = 0$ at all times; we refer to these motions as overshoot \citep[which is discussed in][]{korre_etal_2019}.
Overshoot occurs because the Ledoux boundary is not the location where convective velocity is zero, but rather the location where buoyant acceleration changes sign due to a sign change in the entropy gradient.

The difference between the left and right panels demonstrates that the CZ consumes the SZ.
Overshooting convective motions entrain low-composition material into the CZ where it is homogenized.
This process increases the size of the CZ and repeats over thousands of convective overturn times until the Ledoux and Schwarzschild criteria predict the same convective boundary.
After this entrainment phase, the convective boundary stops moving.
The boundary is stationary because the radiative flux renews the stable temperature gradient; there is no analogous process to reinforce the composition gradient\footnote{Nuclear timescales are generally much longer than dynamical timescales and can be neglected as a source of composition.}.

In Fig.~\ref{fig:profiles}, we visualize vertical profiles in the initial state (left) and evolved state (right).
Shown are the composition $\mu$ (top), the discriminants $\yL$ and $\yS$ (middle), and two important frequencies (bottom): the square \brunt$\,$ frequency $N^2$ and the square convective frequency,
\begin{equation}
f_{\rm{conv}}^2 = \frac{|\vec{u}|^2}{\ell_{\rm{conv}}^2},
\label{eqn:fconv2}
\end{equation}
with $|\vec{u}|$ the horizontally-averaged velocity magnitude and $\ell_{\rm{conv}}$ the depth of the Ledoux-unstable layer.

The composition is initially uniform in the CZ (${z \lesssim 1}$) and RZ ($z \gtrsim 2$), but varies linearly in the SZ ($z \in [1, 2]$).
We have $\yL(z\approx1) = 0$ but $\yS(z\approx2) = 0$.
An unstable boundary layer at the base of the CZ drives the instability and has negative $N^2$.
For $z \gtrsim 1$, we have positive $N^2$, which is larger in the RZ than the SZ.
We found similar results in simulations where $N^2$ was constant across the RZ and SZ.

In the evolved state (right panels), the composition (top) is well-mixed in the CZ and hashed overshoot zone, but \editone{decreases rapidly above the overshoot region.}
We take the height where the horizontally-averaged kinetic energy falls below 10\% of its bulk-CZ value to be the top of the overshoot zone.
\editone{Rare convective events provide turbulent diffusion above the overshoot zone and smooth the profile's transition from its CZ value to its RZ value.}
In this \editone{evolved} state, the Schwarzschild and Ledoux criteria agree upon the location of the convective boundary (middle).

The rate at which the CZ entrains the SZ depends on the stiffness of the radiative-convective interface,
\begin{equation}
\mS = \frac{N^2|_{\rm{RZ}}}{f_{\rm{conv}}^2|_{\rm{CZ}}},
\label{eqn:stiffness}
\end{equation}
which is related to the Richardson number ${\rm{Ri} = \sqrt{\mS}}$.
The time to entrain the SZ is roughly ${\tau_{\rm{entrain}} \sim (\delta h / \ell_{\rm{c}})^2 \rm{R}_{\rho}^{-1} \mS \tau_{\rm{dyn}}}$ \citep[per][eqn.~3]{fuentes_cumming_2020}, where $\delta h$ is the depth of the SZ, $\ell_{\rm{c}}$ is the characteristic convective length scale, $\Ro \in [0, 1]$ is the density ratio \citep[see][eqn.~7]{garaud_2018}, and $\tau_{\rm{dyn}}$ is the dynamical timescale which in our simulation is the convective overturn timescale.
In Fig.~\ref{fig:profiles}, bottom right panel, we have $f_{\rm{conv}}^2|_{\rm{CZ}} \approx 3 \times 10^{-3}$ and $N^2|_{\rm{RZ}} \approx 10^2$, so $\mS \approx 3 \times 10^4$.
Convective boundaries in stars often have $\mS \gtrsim 10^6$, so our simulation is in the same high-$\mS$ regime as stars.
The value of $\mathrm{R}_{\rho}$ can vary greatly throughout the depth of an SZ in a star; we use $\rm{R}_\rho = 1/10$.
The relevant evoluationary timescale during the main sequence is the nuclear time $\tau_{\rm{nuc}}$.
Since $\tau_{\rm{nuc}}/\tau_{\rm{dyn}} \gg (\delta h/\ell_{\rm{c}})^2 \mS/\Ro$ even for $\mS \sim 10^6$, SZs should be immediately entrained by bordering CZs on the main sequence and during other evolutionary stages in which convection reaches a steady state.
Note that while values of $\Ro \ll 1$ increase $\tau_{\rm{entrain}}$, they also support efficient mixing by ODDC (see Sec.~\ref{sec:conclusions}).

\begin{figure}[t!]
\centering
\includegraphics[width=\columnwidth]{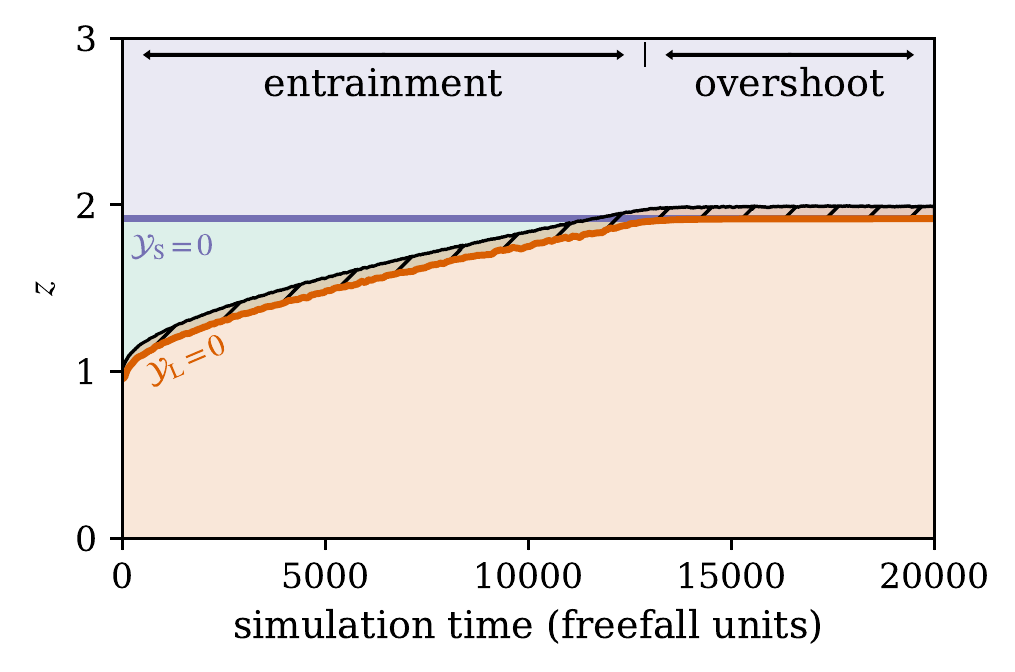}
\caption{
    A Kippenhahn-like diagram of the simulation evolution.
    The $y$-axis is simulation height and the $x$-axis is simulation time.
    The orange line denotes the Ledoux convective boundary ($\yL = 0$); the CZ is below this and is colored orange.
    The purple line denotes the Schwarzschild convective boundary ($\yS = 0$); the RZ is above this and is colored purple.
    The SZ between these boundaries is colored green.
    The black line denotes the top of the overshoot zone, which is hashed.
    The simulation has an ``entrainment phase'', in which the CZ expands, and a pure ``overshoot phase'', in which the convective boundary remains stationary.
\label{fig:kippenhahn}
}
\end{figure}

Finally, in Fig.~\ref{fig:kippenhahn} we display a Kippenhahn-like diagram of the simulation's evolution.
This diagram demonstrates how the CZ, SZ, and RZ boundaries evolve.
The convective boundary measurements are shown as orange ($\yL = 0$) and purple ($\yS = 0$) lines.
The CZ is colored orange and fills the region below the Ledoux boundary, the RZ is colored purple and fills the region above the Schwarzschild boundary, and the SZ is colored green and fills the region between these boundaries.
Convection motions overshoot beyond the Ledoux boundary into a hashed overshoot zone, which we define identically to the one displayed in Fig.~\ref{fig:profiles}.
The top of the overshoot zone (black line) correspond with the edge of the well-mixed region (Fig.~\ref{fig:profiles}, upper right).
While the Schwarzschild and Ledoux boundaries start at different heights, 3D convective mixing causes them to converge.

We briefly note that we performed additional simulations with the same initial stratification as in Fig.~\ref{fig:profiles} (left), but with lower values of $\mS$, higher and lower values of $\Ro$, and less turbulence (lower Reynolds number), and the evolutionary trends described here are present in all simulations.

\section{Conclusions \& Discussion}
\label{sec:conclusions}

In this letter, we present a 3D simulation of a convection zone adjacent to a compositionally stable and weakly thermally unstable region.
This region is stable according to the Ledoux criterion, but unstable according to the Schwarzschild criterion.
Overshooting convective motions entrain the entire Schwarzschild-unstable region until the Schwarzschild and Ledoux criteria both predict the same boundary of the convection zone.

This simulation demonstrates that the Ledoux criterion \emph{instantaneously} predicts the location of the convective boundary, but the Schwarzschild criterion correctly predicts its location on evolutionary timescales (for $t_{\rm{evol}} \gg (\delta h / \ell_{\rm{c}})^2\Ro^{-1}\mS t_{\rm{dyn}}$, see Sec.~\ref{sec:results}).
Our 3D simulation supports the claim that logically consistent implementations of mixing length theory \citep{gabriel_etal_2014, mesa4, mesa5} should have convective boundaries which are Schwarzschild-stable.
E.g., the MESA software instrument's ``convective pre-mixing'' \citep[CPM,][]{mesa5} is consistent with our simulation.
Given our results, the predictions made by 1D stellar evolution calculations should not depend on the choice of stability criterion used if/when convective boundary treatments are properly implemented and $t_{\rm{evol}} \gg (\delta h / \ell_{\rm{c}})^2\Ro^{-1}\mS t_{\rm{dyn}}$.

In stars, SZs should often be unstable to oscillatory double-diffusive convection (ODDC).
\citet{mirouh_etal_2012} show that convective layers often emerge from ODDC, and thus mix composition gradients more rapidly than entrainment alone; ODDC is discussed thoroughly in \citet{garaud_2018}.
\citet{moore_garaud_2016} apply ODDC to SZs which form outside core convection zones in main sequence stars, and their results suggest that ODDC should rapidly mix these regions.
Our simulation demonstrates that entrainment should prevent SZs from ever forming at convective boundaries.

For stages in stellar evolution where ${t_{\rm{evol}} \sim (\delta h / \ell_{\rm{c}})^2\Ro^{-1}\mS t_{\rm{dyn}}}$, time-dependent convection \citep[TDC,][]{tdc_1986} implementations can be used to improve accuracy.
These implementations should include time-dependent entrainment models to properly advance convective boundaries \citep[e.g.,][]{turner_1968, fuentes_cumming_2020}.

\citet{anders_etal_2022} showed convective motions can extend significantly into the radiative zones of stars via ``penetrative convection.''
In this work, we used parameters which do not have significant penetration.
This can be seen in the right panels of Fig.~\ref{fig:profiles}, because the composition is well-mixed above the convective boundary, but the thermal structure is not.

We assume that the radiative conductivity and $\gradrad$ do not depend on $\mu$ for simplicity.
The nonlinear feedback between these effects should be studied in future work, but we expect that our conclusions are robust.

In summary, we find that the Ledoux criterion provides the instantaneous location of the convective boundary, and the Schwarzschild criterion provides the location of the convective boundary in a statistically stationary state; in this final state, the Ledoux and Schwarzschild criteria agree.

\begin{acknowledgments}
We thank Anne Thoul, Dominic Bowman, Jared Goldberg, Tim Cunningham, Falk Herwig, and Kyle Augustson for useful discussions which helped improve our understanding.
    \editone{We thank the anonymous referee for their constructive feedback which improved the clarity of this manuscript.}
EHA is funded as a CIERA Postdoctoral fellow and would like to thank CIERA and Northwestern University. 
The Flatiron Institute is supported by the Simons Foundation.
DL and IGC are supported in part by NASA HTMS grant 80NSSC20K1280.
AEF acknowledges support from NSF Grant Nos.~AST-1814327 and AST-1908338. 
IGC acknowledges the support of the University of Colorado’s George Ellery Hale Graduate Student Fellowship.
MJ acknowledges support from the Barry M. Lasker Data Science Fellowship awarded by the Space Telescope Science Institute.
JRF acknowledges support from a McGill Space Institute (MSI) Fellowship.
This research was supported in part by the National Science Foundation under Grant No. PHY-1748958, and we acknowledge the hospitality of KITP during the Probes of Transport in Stars Program.
Computations were conducted with support from the NASA High End Computing (HEC) Program through the NASA Advanced Supercomputing (NAS) Division at Ames Research Center on Pleiades with allocation GID s2276.
\end{acknowledgments}

\newpage
\appendix

\section{Model \& Initial Conditions}
\label{app:model}
We study incompressible, Boussinesq convection \editone{in which we evolve both} temperature $T$ and concentration $\mu$.
The nondimensional equations of motion are
\begin{align}
    &\grad\dot\vec{u} = 0 
        \label{eqn:incompressible} \\
    &\partial_t \vec{u} + \vec{u}\dot\grad\vec{u} + \grad \varpi = \left(T - \frac{\mu}{\Ro}\right) \hat{z} + \frac{\Pran}{\rm{Pe}}\grad^2 \vec{u}
        \label{eqn:momentum}, \\
    &\partial_t T + \vec{u}\dot(\grad T - \hat{z}\,\partial_z T_{\rm{ad}})   = \grad\dot[\kappa_{T,0} \grad \overline{T}] +  \frac{1}{\rm{Pe}}\grad^2 T'
        \label{eqn:temperature},\\
    &\partial_t \mu + \vec{u}\dot\grad \mu = \frac{\tau_0}{\rm{Pe}}\grad^2\bar{\mu} + \frac{\tau}{\rm{Pe}}\grad^2 \mu',
        \label{eqn:composition}
\end{align}
where $\vec{u}$ is velocity.
Overbars denote horizontal averages and primes denote fluctuations around that average such that $T = \bar{T} + T'$.
The adiabatic temperature gradient is $\partial_z T_{\rm{ad}}$ and the nondimensional control parameters are
\begin{equation}
\begin{split}
    &\mathrm{Pe} = \frac{u_{\rm{ff}} h_{\rm{conv}}}{\kappa_T},\qquad
    \Ro = \frac{|\alpha|\Delta T}{|\beta|\Delta \mu},\qquad\\
    &\Pran = \frac{\nu}{\kappa_T},\qquad
    \tau = \frac{\kappa_\mu}{\kappa_T},\qquad
\end{split}
\end{equation}
where the nondimensional freefall velocity is $u_{\mathrm{ff}} = \sqrt{|\alpha|g h_{\rm{conv}}\Delta T}$ (with gravitational acceleration $g$), $h_{\rm{conv}}$ is the initial depth of the convection zone, \editone{the constant} $\Delta \mu$ is the \editone{initial} composition change across the Ledoux stable region, \editone{the constant} $\Delta T = h_{\rm{conv}}(\partial_z T_{\rm{rad}} - \partial_z T_{\rm{ad}})$ is the \editone{initial} superadiabatic temperature scale of the convection zone, \editone{$\alpha \equiv (\partial \ln \rho / \partial T)|_{\mu}$ and $\beta \equiv (\partial \ln \rho / \partial \mu)|_{T}$} are respectively the coefficients of expansion for $T$ and $\mu$, the viscosity is $\nu$, $\kappa_T$ is the thermal diffusivity, and $\kappa_\mu$ is the compositional diffusivity.
\editone{In stellar structure modeling, $\Ro = |N_{\rm{structure}}^2/N_{\rm{composition}}^2|$ is the ratio of respectively the thermal and compositional components of the \brunt$\,$frequency as measured in a semiconvection zone or thermohaline zone.}
Eqns.~\ref{eqn:incompressible}-\ref{eqn:composition} are identical to Eqns.~2-5 in \citet{garaud_2018}, except we modify the diffusion coefficients acting on $\bar{T}$ ($\kappa_{T,0}$) and $\bar{\mu}$ ($\tau_0$).
By doing this, we keep the turbulence (Pe) uniform throughout the domain while also allowing the radiative temperature gradient $\partial_z T_{\rm{rad}} = -\rm{Flux}/\kappa_{T,0}$ to vary with height.
We furthermore reduce diffusion on $\bar{\mu}$ to ensure its evolution is due to advection.

We define the Ledoux and Schwarzschild discriminants
\begin{equation}
    \yS = \left(\frac{\partial T}{\partial z}\right)_{\rm{rad}} - \left(\frac{\partial T}{\partial z}\right)_{\rm{ad}},\,\,
    \yL = \yS - \Ro^{-1} \frac{\partial \mu}{\partial z},
\end{equation}
and in this nondimensional system the square {\brunt} frequency is $N^2 = -\yL$.

We study a three-layer model with $z \in [0, 3]$,
\begin{align}
    &\left(\frac{\partial T}{\partial z}\right)_{\rm{rad}} = 
    \left(\frac{\partial T}{\partial z}\right)_{\rm{ad}} + 
    \begin{cases}
        -1           & z \leq 2 \\
        10\Ro^{-1}     & z > 2
    \end{cases},
    \label{eqn:initial_T}
    \\
    &\frac{\partial \mu_0}{\partial z} = 
    \begin{cases}
        0        & z \leq 1 \\
        -1       & 1 < z \leq 2 \\
        0        & 2 > z
    \end{cases},
    \label{eqn:initial_mu}
\end{align}
We set $(\partial T / \partial z)_{\rm{ad}} = -1 - 10\Ro^{-1}$.
The intial temperature profile has $\partial_zT_0 = \partial_z T_{\rm{rad}}$ everywhere except between $z = [0.1, 1]$ where $\partial_zT_0 = \partial_z T_{\rm{ad}}$.
\editone{Step functions are not well represented in pseudospectral codes, so we use smooth heaviside functions (Eqn.~\ref{eqn:heaviside}) to construct these piecewise functions.}
To obtain $T_0$, we numerically \editone{integrate the smooth} $\partial_z T_0$ profile with $T_0(z=3) = 1$.
To obtain $\mu_0$, we numerically integrate \editone{the smooth} Eqn.~\ref{eqn:initial_mu} with $\mu_0(z=0) = 0$.

For boundary conditions, we hold ${\partial_z T = \partial_z T_0}$ at $z = 0$, $T = T_0$ at $z = 3$, and we set ${\partial_z \mu = \hat{z}\dot\vec{u} = \hat{x}\dot\partial_z\vec{u} = \hat{y}\dot\partial_z\vec{u}(z=0) = \hat{y}\dot\partial_z\vec{u}(z=3) = 0}$ at $z = [0,3]$.
The simulation in this work uses $\rm{Pe} = 3.2 \times 10^3$, $\Ro^{-1} = 10$, $\rm{Pr} = \tau = 0.5$, $\tau_0 = 1.5 \times 10^{-3}$, and ${\kappa_{T,0} = \rm{Pe}^{-1}[(\partial T/\partial z)_{\rm{rad}}|_{z=0}] / (\partial T/\partial z)_{\rm{rad}}}$
\editone{
    The convective cores of main sequence stars with ${2M_\odot \lesssim M_* \lesssim 10M_\odot}$ have $\rm{Pe} = \mathcal{O}(10^6)$, $\tau \approx \rm{Pr} = \mathcal{O}(10^{-6})$, and stiffnesses of $\mS = \mathcal{O}(10^{6-7})$ (see Jermyn et al.~2022, ``An Atlas of Convection in Main-Sequence Stars'', in prep).
    Our simulation is as turbulent as possible while also achieving the long-term entrainment of the Ledoux boundary, and is qualitatively in the same regime as stars ($\rm{Pe} \gg 1$, $\rm{Pr} < 1$, $\mS \gg 1$).
    Unfortunately, stars are both more turbulent and have stiffer boundaries than can be simulated with current computational resources.
}

\section{Simulation Details \& Data Availability}
\label{app:simulation_details}
We time-evolve equations \ref{eqn:incompressible}-\ref{eqn:composition} using the Dedalus pseudospectral solver \citep[git commit 1339061]{burns_etal_2020} using timestepper SBDF2 \citep{wang_ruuth_2008} and CFL safety factor 0.3.
All variables are represented using a Chebyshev series with 512 terms for $z \in [0, 2.25]$, another Chebyshev series with 64 terms for $z \in [2.25, 3]$, and Fourier series in the periodic $x$ and $y$ directions with 192 terms each.
Our domain spans $x \in [0, L_x]$, $y \in [0, L_y]$, and $z \in [0, L_z]$ with $L_x = L_y = 4$ and $L_z = 3$.
To avoid aliasing errors, we use the 3/2-dealiasing rule in all directions.
To start our simulations, we add random noise temperature perturbations with a magnitude of $10^{-6}$ to the initial temperature field.

Spectral methods with finite coefficient expansions cannot capture true discontinuities.
To approximate discontinuous functions such as Eqns.~\ref{eqn:initial_T} \& \ref{eqn:initial_mu}, we define a smooth Heaviside step function centered at $z = z_0$,
\begin{equation}
H(z; z_0, d_w) = \frac{1}{2}\left(1 + \mathrm{erf}\left[\frac{z - z_0}{d_w}\right]\right).
\label{eqn:heaviside}
\end{equation}
where erf is the error function and we set $d_w = 0.05$.

We produced figures \ref{fig:profiles} and \ref{fig:kippenhahn} using matplotlib \citep{hunter2007, mpl3.3.4}.
We produced figure \ref{fig:dynamics} using plotly \citep{plotly} and matplotlib.
The Python scripts used to run the simulation and to create the figures in this paper are publicly available in a git repository (\url{https://github.com/evanhanders/schwarzschild_or_ledoux}); the data in the figures is available online in a Zenodo repository \citep{supp}.

\bibliographystyle{aasjournal}
\bibliography{biblio}
\end{document}